\documentclass[11pt,twoside]{article}
\usepackage{./asp2014}
\usepackage{color}

\hypersetup{
  colorlinks= black, 
  urlcolor  = blue, 
  linkcolor = black, 
  citecolor = black 
} 

\aspSuppressVolSlug
\resetcounters

\begin{document}

\title{An ngVLA Wide Area AGN Survey}
\author{Allison Kirkpatrick$^{1,2}$, Kirsten Hall$^3$, Kristina Nyland$^4$, Mark Lacy$^{4}$, and Isabella Prandoni$^5$
\affil{$^1$University of Kansas, Lawrence, KS, USA; \email{akirkpatrick@ku.edu}}\affil{$^2$Yale University, New Haven, CT, USA}\affil{$^3$Johns Hopkins University, Baltimore, MD, USA}\affil{$^4$National Radio Astronomy Observatory, Charlottesville, VA, USA}\affil{$^5$INAF - IRA, Via P. Gobetti 101, 40129 Bologna, Italy}}

\paperauthor{Sample~Author1}{Author1Email@email.edu}{ORCID_Or_Blank}{Author1 Institution}{Author1 Department}{City}{State/Province}{Postal Code}{Country}
\paperauthor{Sample~Author2}{Author2Email@email.edu}{ORCID_Or_Blank}{Author2 Institution}{Author2 Department}{City}{State/Province}{Postal Code}{Country}
\paperauthor{Sample~Author3}{Author3Email@email.edu}{ORCID_Or_Blank}{Author3 Institution}{Author3 Department}{City}{State/Province}{Postal Code}{Country}

\begin{abstract}
The next generation Very Large Array (ngVLA) will have unprecedented sensitivities and mapping speeds at $1-8$\,GHz. We discuss how the active galactic nuclei (AGN) community can benefit from a wide-area, medium depth ngVLA survey. We propose a 10 deg$^2$ survey in the Stripe 82 field using the 8\,GHz band with an rms depth of $1\,\mu$Jy~beam$^{-1}$. We will detect $\sim$130,000 galaxies, including radio-quiet AGN out to $z\sim7$. We can measure the luminosity and space density evolution of radio-quiet and radio-loud AGN. We can also measure AGN evolution through clustering of both populations using cross-correlation functions. A wide area ngVLA survey will benefit from existing multiwavelength AGN populations, particularly in the Stripe 82 field, as well as new information from next-generation optical and infrared survey instruments such as LSST and {\it WFIRST}.
\end{abstract}

\section{Introduction}
Likely every massive galaxy has a supermassive black hole at its center, and this black hole is predicted to have a regulatory effect on the buildup of stellar mass due to the tight relationship between stellar mass and black hole mass \citep[e.g.][]{magorrian1998}
On a cosmic scale, the star formation rate density (SFRD) peaks at $z\sim1-3$, and this is an epoch where the black hole accretion rate density peaks as well \citep{madau2014}, although how much of this stellar and black hole mass assembly is occurring within the same galaxies is currently unknown. Probing stellar and black hole mass assembly within similar populations requires wide-area deep surveys, due to the necessity of detecting faint star formation and black hole accretion.
Futhermore, studying the large-scale environments and clustering properties of active galactic nuclei (AGN; the phase in which the central black hole is actively accreting) can elucidate their co-evolution with their host galaxies as well as their greater role in galaxy evolution.

The next generation of radio telescopes (ngVLA and SKA) will enable very deep radio surveys, allowing astronomers to detect {\it all} AGN, not only the tiny fraction of classical radio-loud AGN, where the radio emission is due to large scale jets. The
current picture of radio emission emanating from AGN is that there are two populations: radiatively efficient AGN (quasars/Seyferts) and radiatively inefficient AGN.

Radiatively inefficient AGN (accretion rates <1\%) are mainly radio-loud (RL), and in the local universe are typically associated
with massive early-type galaxies. Radio-mode feedback (so-called ``maintenance mode'') is observed in these massive red systems. The enhanced radio emission can be used to calculate the mechanical energy output into the intergalactic medium \citep{willott1999}, which is observed to heat the centers of clusters and likely serves as the key mechanism that prevents star formation in these systems. A common approach to constraining the importance of radio-mode feedback has been to measure the evolution of the luminosity density of RL AGN, assuming that the evolving luminosity functions directly trace the evolving kinetic energy output \citep{smolcic2017}.
Currently unknown is how much maintenance mode feedback could arise from radio-quiet (RQ) AGN through sub-kpc jets. This is particularly relevant at $z>2$, where the RQ AGN luminosity function dominates over RL AGN \citep{smolcic2017}. 

Radiatively efficient AGN are a more enigmatic population. They are mainly RQ ($L_{1.4\,GHz}<10^{24}\,$W/Hz) quasars and Seyferts (only 10\% of quasars/Seyferts are RL), so the radio emission may come entirely from star formation. Typically, they are AGN identified at other wavelengths and then followed up with radio observations.
However, 30\% of radio-selected RQ AGN have a radio excess (that is a brighter radio flux than would be expected from the standard far-IR/radio correlation), raising the question of whether small-scale radio jets can contribute to the emission of RQ AGN \citep{delvecchio2017,smolcic2017}. Another possibility for the origin of the radio emission in RQ AGN is shocks associated with quasar-driven winds \citep{zakamska2014}. The radio emission is commonly used to calculate the star formation rate (SFR) in RQ AGN, while AGN emission is seen to contaminate other wavelengths. The radio spectral slope in RQ AGN may differ non-negligibly from star forming galaxies (SFGs), so that a correction is required before converting to an SFR. Relatively little is known about the radio emission of the RQ population, due to the necessity of a deep and wide survey, high angular resolution to separate the AGN from the host, and differing AGN selection techniques. 
Here, we outline how ngVLA capabilities can accomplish such a survey in $~$200 hrs, and what the benefit will be.

\section{Radio Surveys: Past and Future}
Building up statistical samples of AGN requires large-area surveys, since AGN are bright but rare. The VLA FIRST survey covered 10,000\,deg$^2$ at 20\,cm at a resolution of $5''$ and an rms of 0.15\,mJy~beam$^{-1}$. From this survey, strong radio emission was used to find a population of extremely dust-reddened quasars, predicted to be an early evolutionary stage before feedback from the AGN blows out the majority of a galaxy's dust \citep{glikman2007,urrutia2009}. The wide area also allowed the detection of a $z\sim6$ radio-loud quasar, demonstrating how rare these objects are \citep{zeimann2011}.

Currently, the VLA Sky Survey (VLASS) is imaging the entire sky visible to the VLA at 2-4\,GHz at a resolution of 2.5$''$ down to an rms of 69\,$\mu$Jy~beam$^{-1}$. This all sky survey is well suited to probing how radio emission might heat the cores of clusters through combination with X-ray all sky surveys from ROSAT, which trace hot gas in clusters.

MIGHTEE is a wide-area survey currently being planned with MeerKAT (the precursor to SKA), which will survey 20\,deg$^2$ down to $\mu$Jy~beam$^{-1}$ sensitivities \citep{jarvis2017}. A survey of this size is designed to detect the environments of low-power radio AGN ($L<10^{24}$\,W\,Hz$^{-1}$), which are experiencing radiatively inefficient accretion, in contrast with RQ AGN \citep{best2012}.

The high sensitivity and angular resolution of the ngVLA will capture emission from AGN in an extinction-free manner at the peak epoch of cosmic assembly. The ngVLA will have 6 receivers spanning 2.4 - 93\,GHz, maximum baselines of up to 1000\,km, and sub-arcsecond resolution. Current and past radio surveys, including MeerKAT, are limited to $\sim$$5''$ resolution, and SKA Phase-I is expected to have resolutions of $0.5-1''$. The unprecedented resolution of the ngVLA will greatly enhance our ability to distinguish high surface brightness AGN emission from jets, and to spatially isolate AGN cores from lower surface brightness emission from star formation, which is critical for identifying low-power radio AGN and for calculating SFR. The bands are wide enough (the 8\,GHz band spans $3.5-12.3$\,GHz) that we will be able to separate the flat spectrum of AGN cores from extended steep spectrum emission from star formation or jets. We describe the details of an ngVLA survey below as well as outlining two science use cases -- mass assembly and clustering.

\subsection{A 10 deg$^2$, 1\,$\mu$Jy ngVLA Survey}
\begin{figure}
\includegraphics[width=2.6in,height=1.8in]{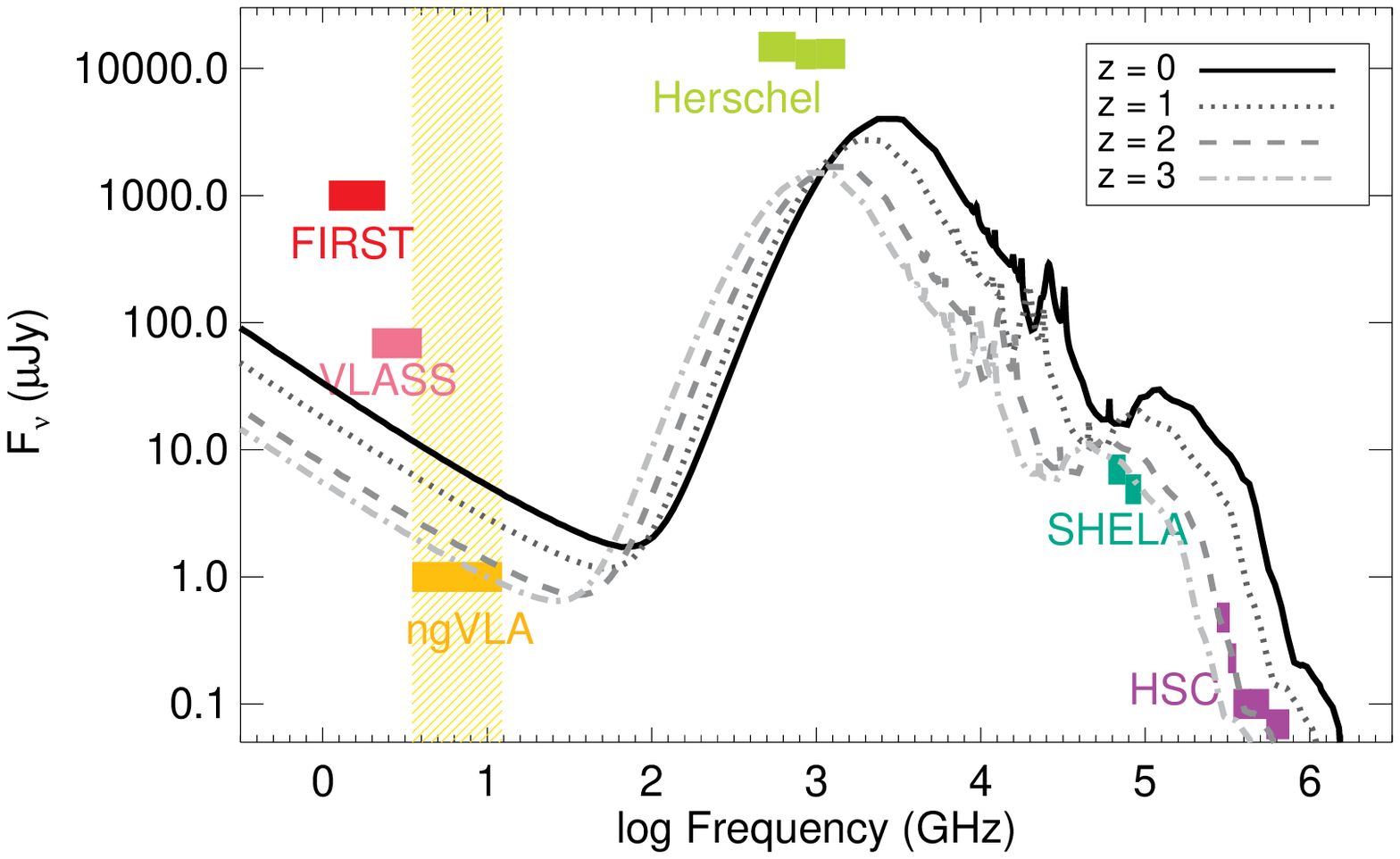}
\includegraphics[width=2.4in,height=2.1in]{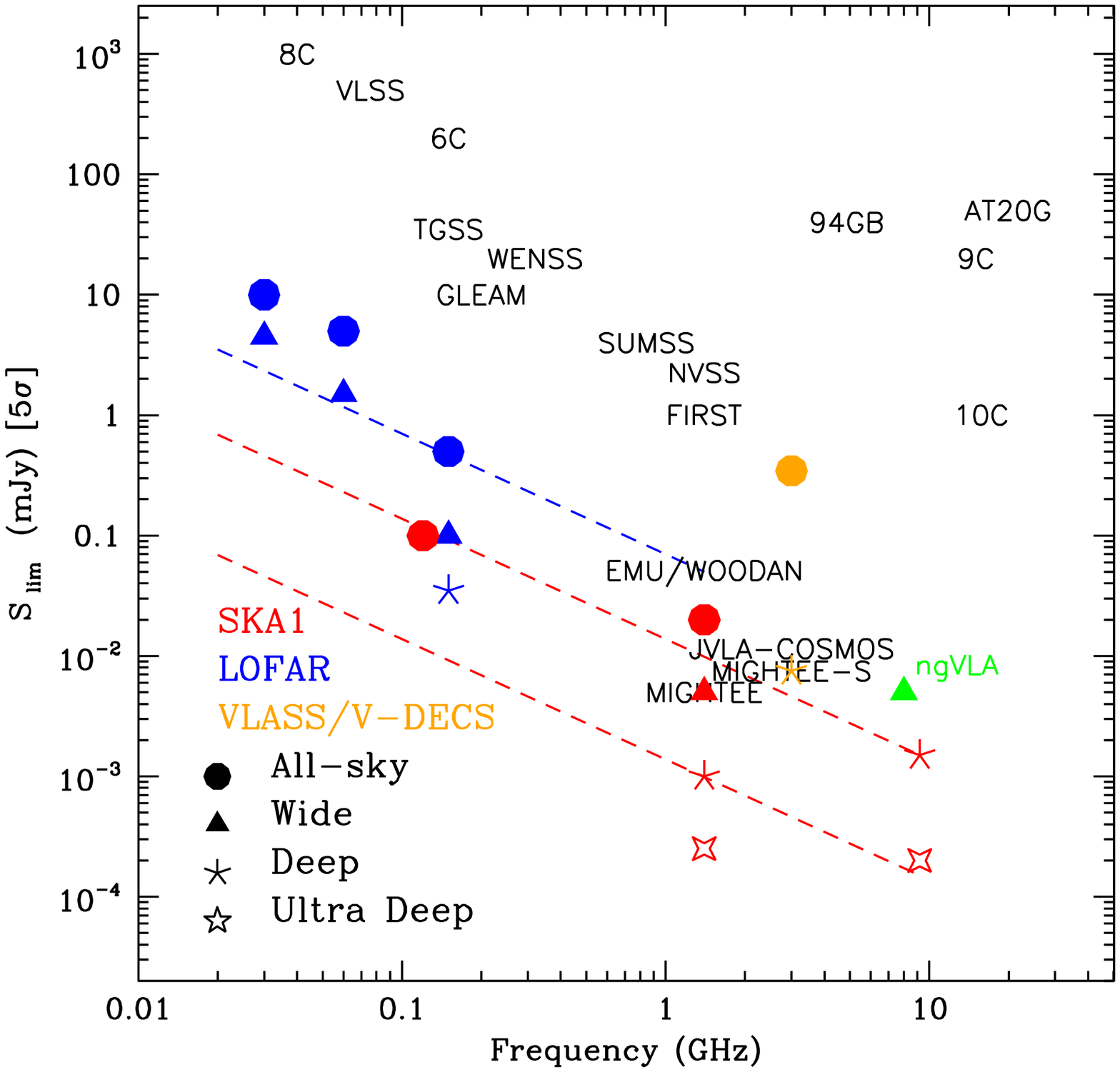}
\caption{(a) {\it Left--} An $L^\ast$ SFG (represented by M82) at $z=0.5$ (solid line), $z=1$ (dotted line), $z=2$ (dashed line), and $z=3$ (dot-dashed line). The ngVLA 8\,GHz band is shaded in orange. The 1\,$\mu$Jy~beam$^{-1}$ rms limit (orange bar) corresponds to an $L^\ast$ galaxy at $z=3$. We illustrate the flux limits of other Stripe 82 surveys: FIRST (red), VLASS (pink), Herschel (chartreuse), SHELA IRAC (aquamarine), and HSC (purple). (b) {\it Right--} $5\sigma$ flux limit vs. frequency for ngVLA in comparison with existing and planned surveys. LOFAR, VLASS, and SKA1
reference surveys are highlighted in blue, orange and red respectively. Different symbols refer to different survey coverage: all-sky (filled circles); wide tiers (filled triangles); deep tiers (asterisks); ultra deep tiers
(starred symbols). The red and blue dashed lines indicate a slope of $\nu^{-1}$ 
for different 1.4\,GHz fluxes.\label{flux_limit}}
\end{figure}

The 8\,GHz receiver is optimal for a wide-field survey due to its  $7.3'$ field of view and sub-arcsecond resolution ($\sim$$0.3''$ with a 36\,km baseline). The dust continuum does not contaminate 8\,GHz observations until $z>8$ making synchrotron emission from  radio-loud AGN easily distinguishable from star forming galaxies (SFGs).
For a 10-square-degree survey at 8~GHz tapered to a resolution of 0.1$^{\prime \prime}$ with a 1$\sigma$ depth of 1\,$\mu$Jy beam$^{-1}$ (given the current ngVLA reference design\footnote{\protect{\href{http://ngvla.nrao.edu/page/refdesign}{http://ngvla.nrao.edu/page/refdesign}}}, we estimate an integration time of $\approx$200~hrs, not including overheads), we will be able to detect $L^\ast$ SFGs at $z<3$ (Figure \ref{flux_limit}a). We calculate $L^\ast$ using an evolving IR luminosity function \citep{gruppioni2013} as this is better constrained than radio $L^\ast$ evolution \citep{padovani2011}. We use M82, a standard SFG, to calculate the predicted radio emission, as the radio to far-IR ratio is consistent for SFGs and RQ AGN \citep{padovani2011}. RL AGN will dominate the survey at $S_\nu>500\mu$Jy \citep{bonzini2013}. Figure \ref{flux_limit}b illustrates the depth of our proposed ngVLA survey in the context of current and future radio surveys.  The ngVLA wide
survey (10 deg$^2$) nicely complements the deeper but narrower SKA surveys
planned for SKA Band 5 (1 and 0.008 deg$^2$ respectively), that will have
similar angular resolution.

\subsection{The Case for Stripe 82}
The Sloan Digital Sky Survey (SDSS, \citealt{york00}; \citealt{abaz09}; \citealt{eise11}; \citealt{daws13}; \citealt{alam15}) Stripe 82 field is a natural choice for a 10\,deg$^2$ legacy survey. 15 deg$^2$ of Stripe 82 has contiguous {\it Chandra} or {\it XMM-Newton} coverage \citep{lamassa2016}, detecting all the brightest X-ray sources ($L_X > 10^{44}$~erg~s$^{-1}$), which are the sources most likely to be radio-loud. Stripe 82 has extensive overlapping multiwavelength coverage, in particular the {\it Spitzer} SPIeS \citep{timlin2016} and SHELA \citep{papovich2016} surveys, {\it Herschel} SPIRE \citep{viero2014}, Subaru HyperSuprimeCam \citep[HSC;][]{miyazaki2018}, as well as VLA FIRST \citep{becker1995} and VLASS \citep{hales2013}. We illustrate the various flux limits in Figure \ref{flux_limit}.

\subsection{Expected Number of Sources}
\label{sec2}
We calculate the expected number of sources in a 1\,$\mu$Jy~beam$^{-1}$, 10 deg$^2$ survey by extrapolating from the JVLA-COSMOS 3\,GHz survey. JVLA-COSMOS covered 1.77\,deg$^2$ with a uniform rms noise of 2.3\,$\mu$Jy~beam$^{-1}$.
The final catalog consists of 10,830 sources detected with S/N$>$5. \citet{novak2017} calculate 1.4\,GHz luminosity functions out to $z\sim 5$
for SFGs\footnote{In this subsection, we follow the convention in \citet{novak2017} and use the term SFG to mean any source where the radio emission arises primarily from star formation. In practice, this can consist of SFGs and RQ AGN.}, and \citet{smolcic2017} does the same for radio excess AGN. Both works assume an $\alpha=-0.7$ to convert from 3\,GHz to 1.4\,GHz. 

The number of sources in a given spherical shell volume is
\begin{equation}
N = \Phi(L,z) \frac{dV_c}{dz\,d\Omega}\,\Omega \,\Delta z\,\Delta\log L
\end{equation}
We extrapolate $\Phi(L,z)$ for AGN and SFGs to the proposed area and flux limit of our survey, setting a detection threshold at S/N$>$3. We use $\alpha=-0.7$ to convert between 1.4 GHz and 8 GHz luminosities. We anticipate observing 112,000 SFGs and 18,000 AGN, for a total of 130,000 sources.

\section{Luminosity Evolution of all AGN out to $z\sim7$}
To date, the best constraints on the evolving radio luminosity function come from deep observations in the COSMOS field \citep{smolcic2017,novak2017} and CDFS field \citep{padovani2011}. Both surveys are hampered by small number statistics, particularly at $z>3$, owing to the relatively small survey areas. Although JVLA-COSMOS covers 2.6\,deg$^2$, only 1.77\,deg$^2$ contains optical/NIR counterparts \citep{laigle2015}, which are needed to measure redshifts and identify AGN.

Our significantly wider field will greatly improve number statistics at high redshift. We will detect $\sim$300 RL AGN at $z>3.3$, an order of magnitude more than JVLA-COSMOS. For SFGs, JVLA-COSMOS reached a limiting redshift of $z\sim5.7$. Our 10\,deg$^2$ survey will detect $\sim$60 SFGs at $z>5.7$, with 4 of those being at $z>6.8$, enabling us to constrain the high-$z$ radio luminosity function. The luminosity density describes how the radio population has evolved over cosmic time. For SFGs, this luminosity density translates directly to a star formation rate density \citep[SFRD; e.g.][]{yun2001,murphy2011}. For RL AGN, it corresponds to a mechanical energy output arising from feedback \citep{willott1999}. We predict the luminosity density evolution observed with an ngVLA survey by assuming SFGs and AGN follow a pure luminosity evolution:
\begin{equation}
\Phi[L,z]=\Phi_0\,\left[\frac{L}{(1+z)^{\alpha_L+z\,\beta_L}}\right]
\end{equation}
We fit for $\alpha_L$ and $\beta_L$ using our extrapolated SFG and AGN luminosity functions (see Section \ref{sec2}). We then integrate $\Phi(L,z)$ to obtain the luminosity density (Figure \ref{lum_func}).

For RL AGN, a wider field may not allow us to detect significantly higher redshift AGN than what have already been found due to the evolution of their space density. \citet{smolcic2017} predict that the space density (and corresponding luminosity density) falls off significantly at $z>3$, hampering their detectability even in wide fields. However, the luminosity and density evolution of RL AGN is still highly uncertain, particularly beyond $z>3$, so our improved number statistics will enable us to more robustly determine the high-$z$ luminosity function.

For radio SFGs, we will detect galaxies up to $z\sim7.5$, beyond the capabilities of any current survey. In JVLA-COSMOS, \citet{novak2017} divide sources into either radio SFGs or radio AGN, based on the ratio of radio emission to the infrared. While the radio emission in SFGs can be attributed to star formation, it does not mean that these galaxies are purely star-forming. They may be selected as RQ AGN, if multiwavelength AGN identification techniques are used, as was done in the CDFS field \citep{padovani2011,bonzini2013}. Our sub-arcsecond resolution will enable us to measure the radio emission in the central kpc and outskirts of every galaxy out to the redshift limit of this survey, cleaning separating the amount of radio emission due to star formation and AGN.

Measuring the SFRD of RQ AGN will determine how much of star formation and black hole growth is truly coeval. \citet{bonzini2013} calculate the fraction of SFGs and RQ AGN as a function of flux density in the VLA CDFS survey, and \citet{padovani2011} estimates that RQ AGN follow a similar space density evolution as SFGs. We use the fractions in \citet{bonzini2013} to estimate how the SFRD curve might break up into the SFR in true SFGs and the SFR in RQ AGN in Figure \ref{lum_func}. 30\% of RQ AGN have a radio emission component likely due to the AGN itself, rather than star formation, so this component can then be converted to a mechanical energy output, as has been done for RL AGN \citep{willott1999}. The sub-arcsecond resolution of an ngVLA survey will allow for a complete picture of mass assembly and energy output in radiatively efficient and inefficient AGN and SFGs.

\articlefigure{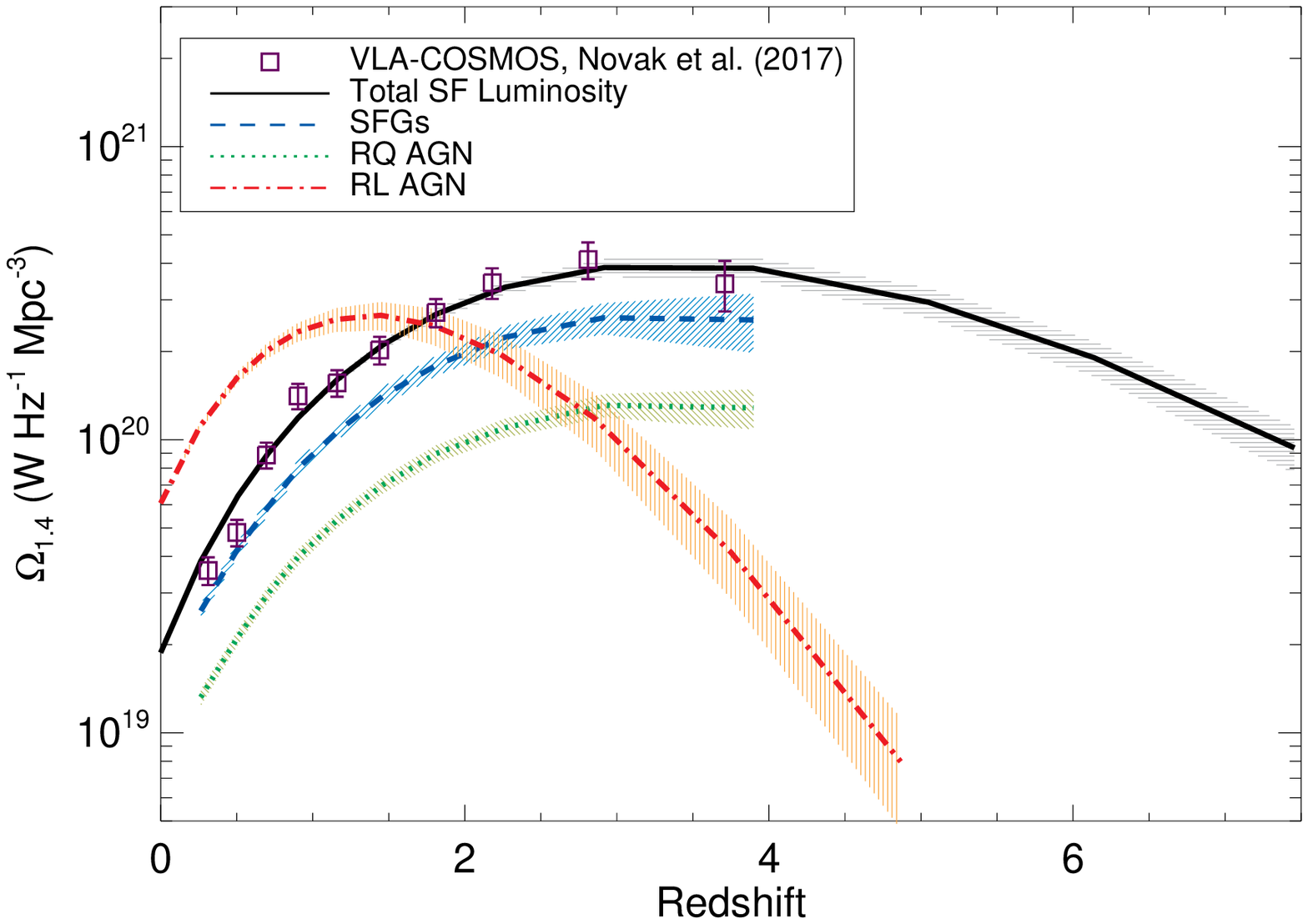}{lum_func}{This figure predicts the radio luminosity functions we would measure with the ngVLA. We have converted 8 GHz to 1.4 GHz to allow easy comparison with the literature. The black curve predicts $\Omega_{1.4}$ from star formation, which we have divided into a component from SFGs (blue) and RQ AGN (green). The current best measurement of the SFG $\Omega_{1.4}$ is shown as the purple squares and comes from VLA-COSMO \citep{novak2017}. The RL AGN $\Omega_{1.4}$ measurements from \citet{smolcic2017} reach the same redshift as illustrated by the red curve. With a wider survey, we will be able to test whether the RL AGN luminosity density really declines so strongly after $z\sim3$.}

\section{Cross Correlation Functions}
Astronomical sky surveys allow for the computation of galaxy correlation functions. 
Of particular use and interest are two point correlation functions: the probability above that of a Poisson distribution of finding two objects within some distance of one another. 
Quantifying auto- and cross-correlation functions of different populations of galaxies yields information about the large-scale structure of the universe, or the way that galaxies cluster together. Analyzing correlation functions as a function of redshift tells about the growth of structure and the evolution of galaxy clusters as a function of cosmic time, since clustering strength of a population should increase as time passes \citep{groth1977}.

The Sloan Digital Sky Survey covers over 35$\%$ of the sky and has spectroscopically catalogued hundreds of thousands of quasars yielding accurate redshifts (out to $z\sim7$) for each. 
Quasars are excellent tracers of large scale structure because they inhabit massive galaxies in small groups and clusters (\citealt{wold00}; \citealt{cold06}; \citealt{trai12}), though they have low spatial density. 
Statistical samples from surveys such as the SDSS and the 2dF QSO Redshift Survey \citep{croo04} enabled the measurement of the quasar auto-correlation function from $z\sim0.5$ (e.g., \citealt{porc04}; \citealt{croo05}) out to $z\ga2.9$ \citep{shen07}.

Quantifying the clustering properties of different types of AGN -- X-ray, radio-loud, radio-quiet, IR bright -- can be informative of any fundamental differences between these populations and how they impact galaxy evolution. 
Clustering studies inform us of the dark matter halo masses in which objects reside as well as how they occupy their dark matter halos (\citealt{peac00}; \citealt{scoc01}; \citealt{berl02}).
For example, \citet{mand09} find that RL AGN live, on average, in halos that are an order of magnitude higher than the RQ population. 
\citet{powe18} cross correlate X-ray selected AGN with 2MASS near-IR galaxies and find that at $z<0.1$ AGN live in group environments, and that obscured AGN tend to occupy denser environments than the unobscured population.

\begin{figure}
\includegraphics[width=5in]{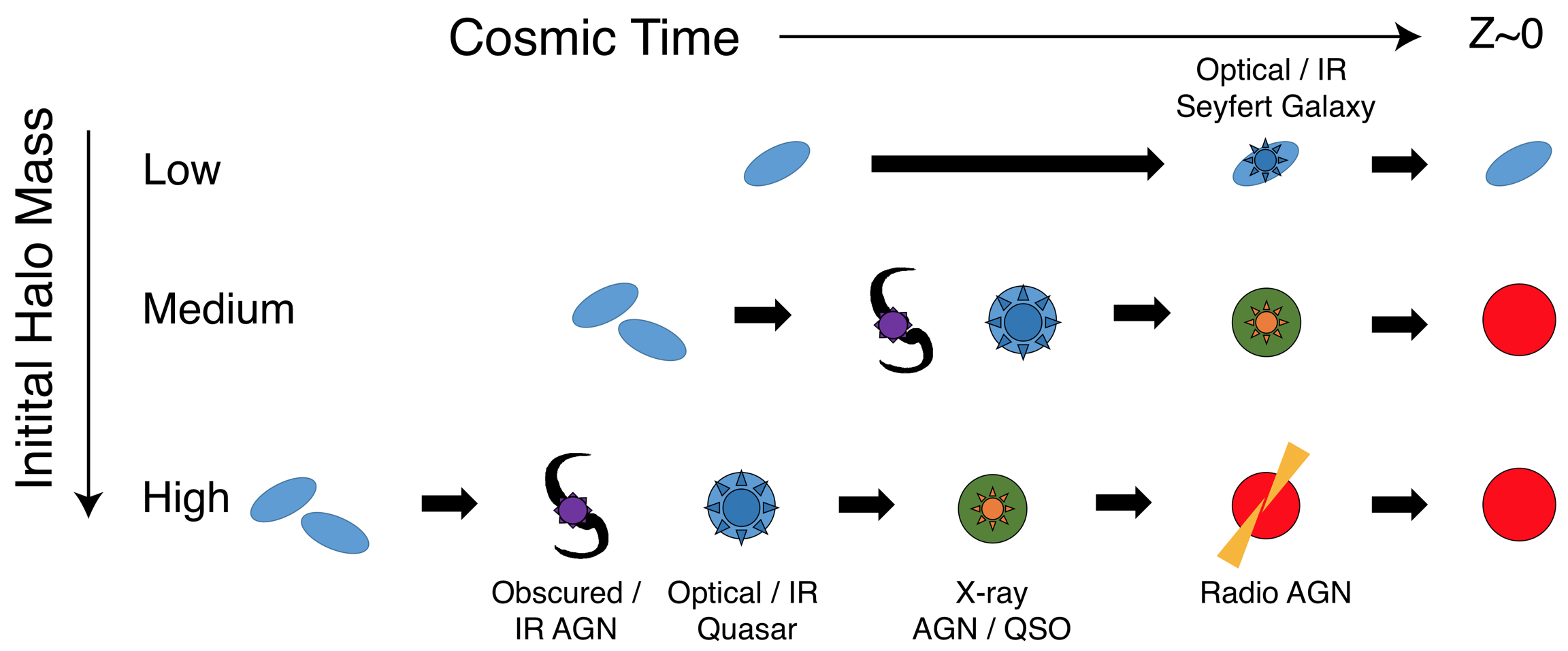}
\caption{Figure adapted from \citet{hickox2009} showing how different wavelength selection of AGN corresponds to a simple evolutionary sequence. The AGN phase begins after halos (dark ovals) reach a critical mass. Rapid black hole growth is initially obscured, followed by an unobscured X-ray phase. After quenching of star formation, red ellipticals are observed as radio AGN, primarily in the centers of massive halos. Measuring clustering of different AGN populations at different redshifts can test this scenario since clustering depends on halo mass.\label{evol}}
\end{figure}

A Stripe 82 ngVLA survey provides a unique opportunity to 
test the evolution of AGN, their host galaxies, and their clustered environments. \citet{hickox2009} measure the clustering and accretion rates of multiwavelength AGN at 
$0.25<z<0.8$ and find that RL AGN are the most strongly clustered, while IR AGN are the least strongly clustered. 
They propose an evolutionary sequence that we illustrate with Figure \ref{evol}. With an ngVLA survey, we will be able 
to measure the clustering of RL AGN, red quasars, and X-ray luminous AGN out to $z\sim4$ (depending on the RL AGN 
space density), for the first time measuring how the clustering of the different populations evolves over cosmic time. 
The proposed survey will even be useful in quantifying physical properties of the RQ population by cross-correlating the survey map with a catalog of known RQ quasars, which by the time of the ngVLA survey will be much deeper than what we have currently as a product of future optical surveys.
Our survey will be able to test at what halo mass the quasar phase occurs, and whether immediately after the 
IR/optical (currently measured with WISE and {\it Spitzer}/IRAC) AGN phase, the galaxy evolves into a RL AGN, producing maintenance mode feedback \citep{hickox2009}.

\section{Summary and Future Directions}
A 10\,deg$^2$, 1\,$\mu$Jy~beam$^{-1}$ ngVLA survey will detect 130,000 galaxies in $\sim$200 hrs, allowing us to characterize the luminosity function of SFGs, RQ AGN, and RL AGN to $z\sim7$. Existing X-ray, optical, and IR counterparts yield photo-$z$s, useful for calculating the clustering of multiwavelength AGN populations. However, by the time the ngVLA begins operations (2030s), 
we will have much deeper optical data than current SDSS. We will have robust photo-z measurements from deep surveys such as HSC and LSST that would be better matched to the increased sensitivity of the ngVLA. LSST will help with AGN identification/characterization through optical variability measurements \citep[e.g.,][]{choi2014}. {\it WFIRST} will provide sub-arcsecond resolution IR imaging that would be well-matched to that of the ngVLA and trace the rest-frame optical emission of high-$z$ galaxies, allowing us to separate emission in the central kpc of galaxies, which should be dominated by the AGN, from the outskirts. eROSITA will provide the first all sky X-ray survey up to 10\,keV, enabling mapping of the hot gas in clusters heated by AGN feedback. Potentially {\it Lynx} will provide sub-arcsecond X-ray observations capable of directly imaging supermassive black holes in the early universe. By combining next generation telescopes with the ngVLA, we will uncover the hidden history of black hole growth through cosmic time.

\acknowledgements The authors thank Jeyhan Kartaltepe for the use of Figure 3, which she adapted from \citet{hickox2009} and Vernesa Smol{\v c}i{\'c} for helpful conversations.

\end{document}